%% file: No_Time_to_Collapse__Unlocking_Robustness_and_Multiplexed_Capacity_in_Frozen_Audio_Watermarkers.tex
\documentclass[letterpaper]{article} 
\usepackage[preprint]{aaai2027}  
\usepackage[hyphens]{url}  
\usepackage{booktabs}
\usepackage{tabularx}
\usepackage{array}
\usepackage{graphicx} 
\usepackage{natbib}  
\usepackage{caption} 
\usepackage{amsmath}   
\usepackage{amssymb}   

\title{No Time to Collapse: Unlocking Robustness and Multiplexed Capacity in
Frozen Audio Watermarkers}

\author{
    Xuanye Wang\equalcontrib\textsuperscript{\rm 2},
    Linxi Li\equalcontrib\textsuperscript{\rm 1,\rm 2},
    Yechen Wang\textsuperscript{\rm 2},
    Liwei Jin\textsuperscript{\rm 2},\\
    Qianwei Guo\textsuperscript{\rm 2},
    Carsten Maple\textsuperscript{\rm 1}
}

\affiliations{
    \textsuperscript{\rm 1}University of Warwick, Coventry, United Kingdom\\
    \textsuperscript{\rm 2}OfSpectrum, Inc., Los Angeles, CA, USA
}

\begin{document}
\maketitle

\begin{abstract}
Modern neural audio watermarking systems typically embed a message repeatedly across time and then 
\emph{collapse} the resulting temporal evidence into a single payload using averaging, voting, or another fixed aggregation rule. We argue that this temporal collapse limits both robustness and the recovery of multiple payloads, and that the limitation can be addressed without retraining the underlying watermarker. We freeze a pretrained watermarker's
encoder and detector and train \emph{only} a low-latency Conformer-based decoder. The decoder consumes the detector's temporal soft outputs, which a system-specific adapter pools into a sequence of window-level representations, and predicts the embedded message. On three frozen watermarkers---AURA,
AudioSeal, and WavMark---the learned decoder improves recovery of attacked
messages and yields higher detection AUROC point estimates on all three.
Under controlled full- and partial-coverage multiplexing, it improves
joint-exact recovery of two
alternating payload words by $9.7$--$48.0$, $6.9$--$17.3$, and
$8.2$--$14.2$ percentage points, respectively, under one to three chained
attacks on feasible clips.
\end{abstract}


\section{Introduction}
\label{intro}
Neural audio watermarking embeds an imperceptible, recoverable message into a
waveform, and has become a practical tool for provenance and attribution of
synthetic speech. Recent systems such as AudioSeal
\citep{audioseal} and WavMark \citep{wavmark} train an encoder--detector pair
end to end: the encoder hides a $k$-bit message, and the detector produces
temporal soft outputs about the payload, per window for some systems and per
sample for others. To return a single message, this temporal evidence is then
\emph{collapsed} across time by a fixed reduction such as an average, a
majority vote, or an error-correcting decode.

We argue that this time-collapsing aggregation is a decisive bottleneck. A fixed reduction is chosen in advance and
cannot adapt to the attack. Whether it averages, votes, or gates on a hard
pattern check, it applies one rule to whatever evidence survives, and it
cannot learn context-dependent weightings of the surviving windows. Moreover, without an aggregation mechanism that can reliably disentangle evidence across time, the released decoding interfaces we evaluate return a single payload per utterance~\cite{audioseal,wavmark}. When multiple messages are embedded and decoding is forced under the same fixed aggregation rule, their evidence can be conflated, leading to ambiguous or unstable predictions. Both limitations are properties of the \emph{decoder} rather than the encoder, so where usable temporal evidence survives they can be removed without retraining the watermarker at all.

In order to address these limitations, we replace the fixed reduction with a small learned sequence model, a
Conformer-based model \citep{conformer} that consumes the frozen detector's temporal soft
outputs, pooled by the adapter into a window sequence, and emits the message.
The encoder and detector stay frozen and no new embedding model is trained. We
consider two settings. In the \emph{\textbf{single-message}} setting, one
message is embedded across the entire utterance, leaving the embedded signal
unchanged, and the decoder improves recovery by combining the temporal evidence
more effectively. In the \emph{\textbf{multiplexing}} setting, two messages are embedded in alternating time slots using the same frozen embedding primitive at unchanged per-slot strength. Our Conformer disentangles the resulting temporal evidence to recover both messages, increasing effective payload capacity without retraining the watermarker.

We apply the same decoder architecture and frozen-base procedure to three watermarkers: our independent reimplementation of AURA \cite{aura}, and the open-source AudioSeal \cite{audioseal} and WavMark \cite{wavmark}.

The contributions of this paper can be summarized as follows:
\begin{itemize}
    \item A \emph{watermarker-agnostic} learned decoder for watermarkers that
    repeat a payload over time and expose temporal soft outputs (one soft-read
    adapter per system and one trained decoder per system and setting): a Conformer trained only on
    the frozen watermarker's temporal soft outputs, replacing its native
    time-collapsing aggregation without touching the encoder or detector.
    \item Evidence across \emph{three} watermarkers that a learned decoder
    improves attacked payload recovery, and also yields higher detection
    AUROC point estimates on all three.
    \item In the \emph{\textbf{multiplexing}} setting, the Conformer recovers
    two alternating payloads from one clip, improving joint-exact recovery over
    grid-aware fixed baselines on all three watermarkers under full and partial
    coverage.

\item A controlled, operating-point-independent evaluation protocol: exact
recovery is scored on known watermarked clips, independently of the presence
decision, and detection is assessed separately by AUROC. All comparisons use held-out data with paired significance analysis.
\end{itemize}

\section{Related Work}

\paragraph{Audio watermarking.}
Classical audio watermarking hides data with hand-designed waveform or
spectral transforms---spread-spectrum embedding \citep{kirovski}, early
data-hiding transforms \citep{bender}, psychoacoustic masking
\citep{swanson}, and quantization index modulation \citep{qim}---under an
explicit trade-off among imperceptibility, capacity, and robustness. More
recent systems jointly train neural embedders and detectors
\citep{pavlovic}. Representative examples include AudioSeal
\citep{audioseal}, which supports localized watermark detection; WavMark
\citep{wavmark}, which uses an invertible chunk-based architecture; and AURA
\citep{aura}, which targets robust high-capacity watermarking. Other systems include
SilentCipher \citep{silentcipher}, MaskMark \citep{maskmark},
re-recording-resilient DeAR \citep{dear}, cross-attention conditioning
\citep{xattnmark}, dual localization/payload embedding \citep{ideaw},
frame-wise generation within TTS \citep{traceablespeech}, and repeated
frequency-domain marks against voice cloning \citep{timbrewatermarking}. Their
robustness is increasingly evaluated using standardized benchmarks
\citep{audiomarkbench,rawbench}.

\paragraph{Temporal organization and native decoding.}
Although watermark capacity has a classical information-theoretic treatment
\citep{moulin}, we focus on the temporal organization of payload evidence.
The released WavMark implementation we use encodes each 1-s chunk with a
32-bit codeword consisting of a 16-bit synchronization pattern and a 16-bit
payload, while our AURA reimplementation assigns 32-bit codewords to
non-overlapping 2-s embedding regions.
These blockwise designs provide disjoint temporal carriers. AudioSeal instead
embeds a 16-bit payload with overlapping temporal support; when different
payloads occupy adjacent 1-s intervals, their watermark evidence can mix near
the boundary. Related work addresses temporal cropping through repeated
embedding and sliding-window recovery \citep{msequence}, while WAKE supports
multiple embeddings and variable-length watermarks \citep{wake}.

Despite these differences, the decoders we evaluate ultimately reduce temporal
evidence to a single payload using fixed rules. AudioSeal averages
per-sample message logits over the clip and aggregates watermark presence
separately. WavMark averages chunk predictions that pass its synchronization-%
pattern check, while our AURA reimplementation error-corrects sliding-window
predictions before combining them by majority vote. WavMark further notes that its chunks may be
combined across a stream to trade capacity for robustness
\citep{wavmark}. In all three systems, however, the temporal reduction is
hand-specified and produces only one clip-level payload.

\paragraph{Learned aggregation and sequence models.} The Conformer
\citep{conformer} interleaves convolution and self-attention and is a strong
sequence encoder for speech, building on Transformer attention
\citep{transformer}. Learned attention pooling has also been used to replace
fixed instance aggregation in multiple-instance learning
\citep{attentionmil}. We use the Conformer not for recognition but as the
\emph{aggregator} that maps a sequence of per-window detector outputs to a
message, so that the decoder can weight evidence by reliability and localize
the payload in time instead of averaging blindly. Recent decoder-centric work
also improves watermark recovery under self-vocoding distortions
\citep{selfvocoding}; our setting instead targets aggregation of temporal
outputs from frozen detectors and separation of multiple payloads.

\section{Method}

\begin{figure*}[t]
  \centering
  \includegraphics[width=0.90\textwidth]{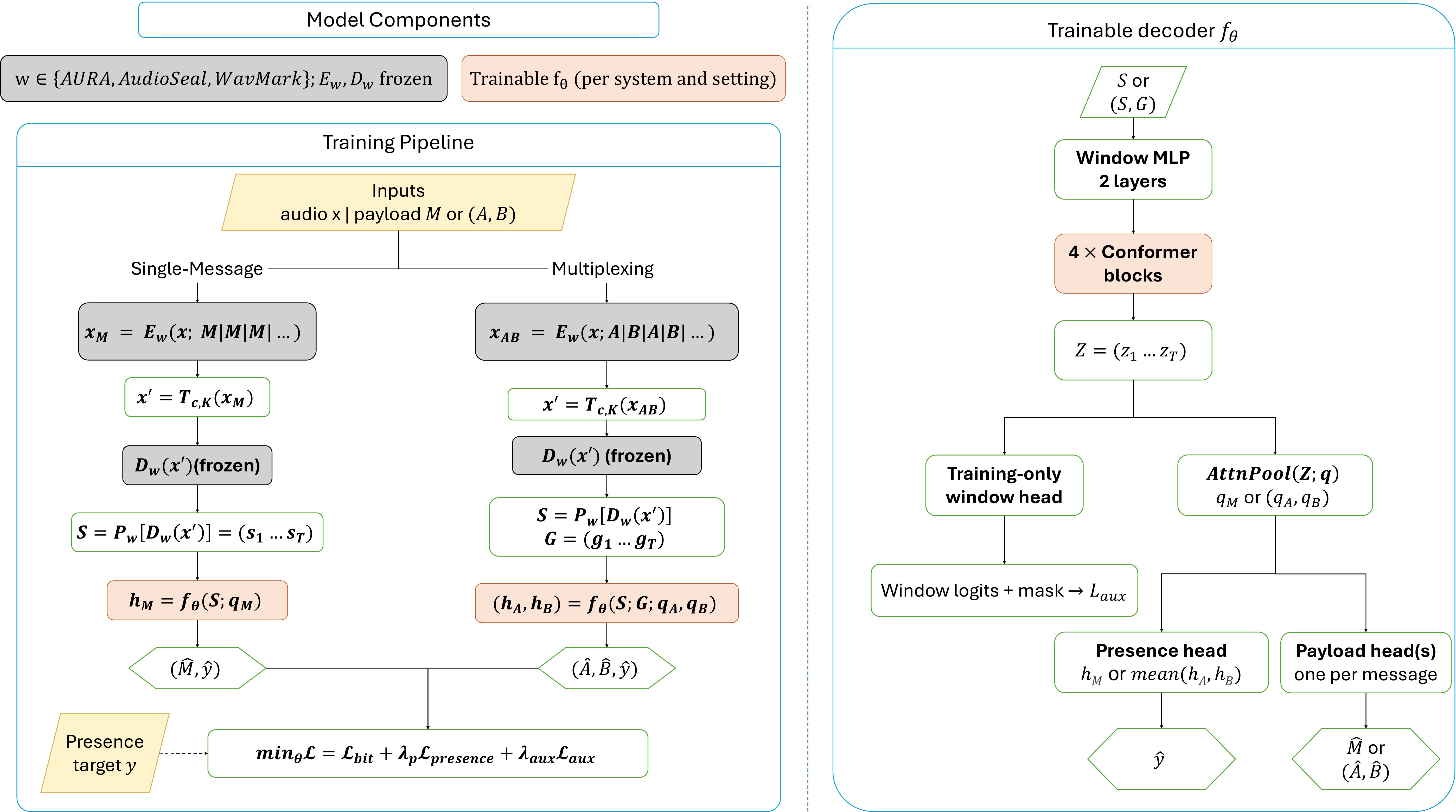}
  \caption{Frozen-base training pipeline. For watermarker $w$, encoder $E_w$ and
detector $D_w$ are frozen; only $f_\theta$ is trained. $E_w$ embeds payload $M$
across audio $x$ (giving $x_M$) or $A,B$ in alternating slots (giving
$x_{AB}$); the attack chain $T_{c,K}$ produces $x'$; $D_w$ emits temporal soft
outputs; adapter $P_w$ pools them into windows $S=(s_1,\ldots,s_T)$, with grid
flags $G=(g_1,\ldots,g_T)$ in multiplex mode. Inside $f_\theta$, a two-layer
window MLP and four Conformer blocks give $Z=(z_1,\ldots,z_T)$; attention
pooling with $q_M$ yields $h_M$, or with $(q_A,q_B)$ yields $(h_A,h_B)$;
payload heads predict $\widehat{M}$ or $(\widehat{A},\widehat{B})$; a presence
head predicts $\widehat{y}$ from $h_M$ or $\operatorname{mean}(h_A,h_B)$; and a
training-only window head produces masked per-window logits for
$\mathcal{L}_{\mathrm{aux}}$. The objective is
$\mathcal{L}_{\mathrm{bit}}+\lambda_p\mathcal{L}_{\mathrm{presence}}
+\lambda_{\mathrm{aux}}\mathcal{L}_{\mathrm{aux}}$; clean negatives ($y{=}0$)
bypass $E_w$ and supervise only $\mathcal{L}_{\mathrm{presence}}$.}

  \label{fig:method-overview}
\end{figure*}

Figure~\ref{fig:method-overview} summarizes the single-message and multiplex
training paths and the shared trainable decoder architecture.

\paragraph{The Time-Collapse Bottleneck.}
A frozen watermarker produces soft detector outputs over time. A
watermarker-specific adapter converts them into an ordered sequence of
window-level vectors
$\mathbf{S}=(\mathbf{s}_1,\ldots,\mathbf{s}_T)$, where
$\mathbf{s}_t\in\mathbb{R}^{d}$. Native decoding applies a predefined
reduction $g(\mathbf{S})$---such as averaging, voting, or pattern-gated
aggregation---to obtain a clip-level payload. Although such rules may reject
low-confidence windows, they cannot learn context-dependent window
reliability. They also typically collapse the sequence to a single payload,
discarding which temporal slots produced the evidence. We replace this
decoder-side reduction while keeping the watermark encoder and detector
frozen.

\paragraph{Learned Sequence Decoder.}
We train a Conformer $f_\theta$ on the window sequence. It uses one
attention-pooling query per target message; each query produces a pooled
representation from which a $k$-bit payload head predicts that message.
Single-message decoding therefore uses one query, whereas two-message decoding
uses two. A separate presence head acts on the single pooled representation,
or their mean in multiplex mode, and produces one clip-level presence logit.
Each watermarker has its own adapter and separately trained decoder instance,
but the decoder architecture is shared. Payload probabilities are binarized
at $0.5$, while the presence output supplies the score used for detection
AUROC and the supplementary end-to-end evaluation.

During training, an optional auxiliary head predicts payloads from individual
window representations. In the single-message setting, majority-watermarked
windows predict the clip payload. In the multiplex setting, slot-pure,
majority-watermarked windows predict the message assigned to their slot;
boundary-crossing and unwatermarked windows are excluded. This encourages the
model to preserve localized payload evidence before pooling.

\paragraph{Temporal Multiplexing (Capacity).}
To construct training and evaluation examples for the multiplexing setting, we embed $A$ and $B$ in alternating slots of a
fixed \textbf{temporal grid}, $A,B,A,B,\ldots$. The grid is a protocol constant known
to every decoder; only the payload assigned to each slot changes, while the
encoder and embedding strength remain fixed. Each window receives a
three-state flag indicating an $A$ slot, a $B$ slot, or no unambiguous
assignment. Two pooling queries then recover the two messages, yielding
$2k$ distinct payload bits per clip. Grid-aware baselines receive the same slot
information, so the comparison is not driven by privileged knowledge of the
grid; any further differences in presence gating or error-correction placement
are reported explicitly.

\section{Experimental Setup}
\label{sec:experimental-setup}

\subsection{Decoder Architecture}
All three watermarkers' encoders and detectors are frozen; only a watermarker-specific
Conformer decoder is trained. Its soft-read adapter supplies 32 AURA logits
from 2-s windows at 0.1-s stride, 16 AudioSeal message logits average-pooled
over non-overlapping 0.1-s windows, or all 32 WavMark
pattern-plus-payload (16 + 16 bits) scores from non-overlapping 1-s chunks. We append the
mean absolute score and mean sigmoid uncertainty to each window. Multiplex
inputs additionally receive a three-state $A$-slot/$B$-slot/boundary flag.

The decoder uses a two-layer window MLP, four Conformer blocks with
$d_{\mathrm{model}}{=}128$, four attention heads, feed-forward width 512,
convolution kernel 15, and dropout $0.1$ (approximately 2\,M parameters).
Single-message decoding uses one learned attention-pooling query followed by
one $k$-bit head. Multiplex decoding uses two queries and two separate
$k$-bit heads, whose outputs are concatenated; both queries attend to the
complete sequence rather than being hard-masked by slot. Overall, the decoder maps an input sequence in
  $\mathbb{R}^{T\times d_{\mathrm{in}}}$ through the window MLP,
  Conformer blocks, attention pooling, and output heads, in that order,
  to a $k$-bit payload in single-message mode or a concatenated $2k$-bit
  payload in multiplex mode, together with a single clip-level presence
  logit.

\subsection{Training}
\paragraph{Data and optimization.}
The source pool contains $1{,}061{,}905$ clips, $96.1\%$ from Emilia
\citep{emilia} and $3.9\%$ from FSD50K \citep{fsd50k}. AudioSeal and WavMark use a deterministic, source-disjoint 90/5/5 split (seed 42); clips are converted to mono at
16\,kHz and truncated or zero-padded to 5\,s for single-message training and
8\,s for multiplex training. Payloads, presence labels, coverage masks, and
attacks are generated online. A clip is watermarked with probability $0.7$,
and half of positive examples use \textbf{partial coverage} (details in Coverage and training attacks below). AdamW \citep{adamw} settings, hardware, and
per-system training budgets are given in the supplementary material.

AudioSeal's single-message decoder is trained on 100k clips and its multiplex
decoder on 40k; WavMark uses 40k for both. We use \texttt{audioseal}~0.2.0 and
\texttt{wavmark}~0.0.3 with their released checkpoints. Single-message
checkpoints are selected by validation exact-message accuracy and multiplex
checkpoints by joint-exact recovery.

For AURA, both Conformer decoders are trained on datasets generated with the
frozen encoder: a 50k-example source-disjoint 70/15/15 split for
single-message, and a separate 303,973-example dataset with alternating 32-bit
payloads for multiplex.

\paragraph{Loss.}
For AudioSeal and WavMark, we optimize
\[
\mathcal{L}=\mathcal{L}_{\mathrm{bit}}
+0.3\mathcal{L}_{\mathrm{presence}}
+0.3\mathcal{L}_{\mathrm{aux}},
\]
where $\mathcal{L}_{\mathrm{bit}}$ is optimized on positive clips, $\mathcal{L}_{\mathrm{presence}}$ on all clips,
and $\mathcal{L}_{\mathrm{aux}}$ on eligible post-Conformer window predictions. In
single-message training, a majority-watermarked window predicts the clip
payload. In multiplex training, it must additionally be slot-pure---contained
within one 2-s slot---and supervises only the payload assigned to that slot.
Boundary and unwatermarked windows remain model inputs but are excluded from
the auxiliary loss. Both selected AURA checkpoints use
$\mathcal{L}_{\mathrm{bit}}+\mathcal{L}_{\mathrm{presence}}
+0.3\mathcal{L}_{\mathrm{aux}}$ with $0.05$ payload-label smoothing.

\paragraph{Coverage and training attacks.}
\label{cover}
\emph{Full coverage} means that the complete clip is watermarked. Contiguous
partial coverage retains one watermarked interval of fraction
$c\sim\mathcal{U}(0.1,0.95)$ at a random offset and replaces the remainder
with clean audio; this is used for AudioSeal and WavMark's single-message training and
WavMark multiplex training. AudioSeal multiplex instead uses four 2-s $ABAB$ slots (8\,s in total); partial coverage keeps one
  $A$ and one $B$ slot (50\%), retaining both payloads. AURA splices watermarked
  and clean 0.5-s tiles with 10-ms crossfades. 
  
  During training, AudioSeal and WavMark independently augment each clip with
  $K\sim\operatorname{Unif}\{1,2,3\}$ distinct attacks from Table \ref{tab:attacks} in both settings; AURA
  uses the same scheme for multiplexing, whereas its reused single-message
  checkpoint was trained with one attack per clip.

\subsection{Evaluation}
All evaluations use the first 11 attacks, which exclude timing warps because
these invalidate the fixed temporal grid used for per-window supervision and
multiplex assignment; only AURA's per-attack evaluation adds the two timing
warps, for 13. Each clip is evaluated under every single attack ($K{=}1$) and
under random chains at $K{=}2,3$, with $K{=}0$ as the clean control.

\begin{table}[t]
\centering
\small
\setlength{\tabcolsep}{3pt}
\begin{tabularx}{\columnwidth}{
    @{}l>{\raggedright\arraybackslash}X@{}}
\toprule
Attack & Operation and sampled setting \\
\midrule
White noise & White Gaussian noise at SNR $s\sim\operatorname{Unif}\{1,\ldots,15\}$\,dB. \\
Pink noise & $1/f$ noise at waveform std $\sigma=0.01$. \\
Low-pass & Attenuate above $f_c\sim\mathcal{U}(3,6)$\,kHz. \\
High-pass & Attenuate below $f_c=500$\,Hz. \\
Smoothing & Moving average, $w\sim\operatorname{Unif}\{2,\ldots,10\}$ samples. \\
Resample & To $f\in\{44.1,24,22.05,16\}\setminus\{f_{\mathrm{op}}\}$\,kHz, then back. \\
Amplitude & Scale $x'\!=g x$, $g\sim\mathcal{U}(0.5,1.1)$. \\
Quantization & Quantize to $q\in\{256,512,1024\}$ levels. \\
Sample dropout & Zero each sample, $p\sim\mathcal{U}(0.005,0.02)$. \\
STFT masking & Mask STFT time/frequency regions, max width 80 bins; adapted
  from SpecAugment \citep{specaugment}. \\
Median filter & Median filter, kernel $w\in\{3,5\}$ samples. \\
\midrule
Time-stretch$^\dagger$ & Pitch-preserving stretch, $r\sim\mathcal{U}(0.92,1.08)$. \\
Speed$^\dagger$ & Change speed (duration and pitch), $r\sim\mathcal{U}(0.90,1.10)$. \\
\bottomrule
\end{tabularx}
\caption{Signal-level augmentations. Parameters are sampled independently
for each application. $\operatorname{Unif}\{\cdot\}$ denotes discrete uniform
sampling and $\mathcal{U}(a,b)$ continuous uniform sampling.
$^\dagger$ Applied only in AURA's per-attack evaluation.}
\label{tab:attacks}
\end{table}

  \paragraph{Evaluation sets.}
  Single-message test sets contain 5,008 four-second AURA clips (3,552
  watermarked), 4,992 AudioSeal clips, and 5,008 WavMark clips; the latter two
  are approximately 50\% positive. During testing, every attack is applied
  separately at $K{=}1$ (Table \ref{tab:per-attack-combined}), whereas
  $K{=}2,3$ use random chains.

\paragraph{Controlled multiplex evaluation.}
  AURA and AudioSeal each contribute 1,504 known-positive 8-s clips, WavMark
  742--763 per cell. Both decoders know the $ABAB$ grid, though not which
  interval partial coverage retains.
  We score only known-positive clips, without detection. Partial rates count
  only feasible clips, where $A$ and $B$ each retain over 0.5\,s of watermarked
  audio within a 1-s interval of a slot; the mask is used only for this filter,
  never as decoder input. Full and partial cells therefore differ in
  composition.

\paragraph{Scoring.}
  Single-message results report \textbf{exact-message accuracy}, i.e., the
  fraction of known-positive clips whose entire payload is recovered
  bit-for-bit. Multiplex results report \textbf{joint-exact recovery}, for which both payloads must match bit-for-bit: two raw 32-bit payloads for AURA and two 16-bit payloads for AudioSeal and WavMark.
  Both metrics are ungated; each
  payload bit is decoded as one when its predicted probability is at least
  $0.5$ (equivalently, at zero logit for AURA).

  Detection is evaluated separately by AUROC over watermarked and clean clips.
  The Conformer uses its learned presence probability, while AudioSeal uses its
  native continuous detector score. Because WavMark's shipped detector returns
  only a binary exact-pattern decision, its baseline AUROC uses the maximum
  signed synchronization-pattern margin over its sliding windows,
  \[
  \max_t \frac{1}{16}\sum_j(2b_j-1)(p_{t,j}-0.5),
  \]
  where $b_j$ is the expected synchronization bit and $p_{t,j}$ its soft score.
  For AURA, the hard-vote baseline uses the constructed window-consistency score
  \[
  \frac{1}{32}\sum_j
  \left|2\,\operatorname{mean}_t\mathbf{1}[z_{t,j}\geq0]-1\right|,
  \]
  where $z_{t,j}$ is the detector logit for bit $j$ in window $t$. These
  constructed scores are used only for threshold-free detection ranking.

  We use $N$ to denote all evaluated clips and $N_{+}$ for the watermarked
  subset; exact-message accuracy uses $N_{+}$, whereas AUROC uses $N$. Paired
  message-recovery differences use continuity-corrected McNemar tests,
  $10^4$-sample paired bootstrap CIs, and Holm--Bonferroni correction; AUROC
  values are reported as point estimates without paired significance tests.

\paragraph{Paired baselines.}
The Conformer and baseline always receive the same attacked waveform and the
same frozen detector, though sampling paths differ by system. Single-message baselines are the native AudioSeal
and WavMark decoders and a hard vote for AURA. Multiplex
baselines receive the same known $A/B$ grid: AURA hard-votes raw slot outputs
by parity, WavMark applies its released chunk decoder within each parity, and
AudioSeal averages the corresponding windows from one continuous full-stream
detector pass. In particular, AudioSeal is not allowed to crop slots and
rerun its detector. Because payload width, sample rate, source set, and
detector geometry differ across systems, we interpret only within-system
Conformer--baseline differences.

\paragraph{Lightweight learned baseline.}
To distinguish Conformer-specific gains from the generic benefit of learned
aggregation, we additionally train a simple BiGRU on the WavMark multiplex
task. It replaces the four Conformer blocks with one bidirectional GRU
(64 hidden units per direction), while retaining the same window MLP, slot
flags, dual-query attention pooling, output heads, losses, training data, and
augmentations. The complete BiGRU decoder has 171k parameters, versus 1.96M
for the Conformer. Both decoders are evaluated on the same paired held-out clips.

\section{Results}

\paragraph{Single-Message: Single-Attack Robustness}
\label{app:per-attack}

Table~\ref{tab:per-attack-combined} shows that the Conformer matches or
outperforms the fixed baseline in 29 of 35 system--attack cases. The six
decreases, all on AudioSeal, are at most $0.7$\,pp and none remains significant
after correction. Gains are generally largest when native recovery is poor but
usable evidence remains, such as AudioSeal under low-pass filtering
($+23.2$\,pp), WavMark under median filtering ($+28.9$\,pp), and AURA under
quantization ($+8.8$\,pp). White noise is the exception: all three systems
remain near floor with gains of at most $0.5$\,pp, consistent with little
recoverable evidence for either decoder.

\input{per_attack_tables}

\paragraph{Single-Message: Combined Attacks Robustness}
Table~\ref{tab:robust} shows that under $K{=}2,3$ attack chains, the Conformer
outperforms the baseline in every exact-message and AUROC cell. Even AudioSeal,
whose single-attack results are mostly tied, gains $+1.4$ to $+4.2$\,pp;
WavMark gains $+8.0$ to $+11.4$\,pp and AURA gains $+3.4$ to $+6.2$\,pp.
Together with Table~\ref{tab:per-attack-combined}, this further suggests that learned
aggregation is most useful when compounded distortions weaken, but do not
eliminate, recoverable evidence.

\begin{table}[t]
\centering
\small
\setlength{\tabcolsep}{3pt}
\begin{tabular}{@{}llccccc@{}}
\toprule
& & \multicolumn{3}{c}{Exact message (\%)} & \multicolumn{2}{c}{AUROC (\%)} \\
\cmidrule(lr){3-5}\cmidrule(lr){6-7}
System & Regime & Conf. & Base. & $\Delta$ & Conf. & Base. \\
\midrule
AudioSeal & Full $K{=}2$    & 58.6 & 54.9 & $+3.7$ & 99.6 & 97.7 \\
          & Full $K{=}3$    & 42.9 & 38.8 & $+4.2$ & 98.8 & 95.5 \\
          & Partial $K{=}2$ & 51.6 & 50.0 & $+1.6$ & 98.2 & 94.8 \\
          & Partial $K{=}3$ & 34.5 & 33.1 & $+1.4$ & 96.5 & 90.9 \\
\midrule
WavMark   & Full $K{=}2$    & 66.0 & 58.0 & \phantom{0}$+8.0$ & 97.0 & 96.2 \\
          & Full $K{=}3$    & 51.0 & 39.7 & $+11.4$ & 93.5 & 92.2 \\
          & Partial $K{=}2$ & 55.5 & 46.3 & \phantom{0}$+9.2$ & 95.7 & 94.6 \\
          & Partial $K{=}3$ & 38.0 & 29.3 & \phantom{0}$+8.7$ & 90.7 & 89.6 \\
\midrule
AURA      & Full $K{=}2$    & 56.8 & 50.6 & \phantom{0}$+6.2$ & 95.2 & 85.1 \\
          & Full $K{=}3$    & 38.5 & 34.2 & \phantom{0}$+4.3$ & 90.1 & 80.3 \\
          & Partial $K{=}2$ & 38.5 & 33.5 & \phantom{0}$+5.0$ & 92.1 & 77.7 \\
          & Partial $K{=}3$ & 24.8 & 21.4 & \phantom{0}$+3.4$ & 86.4 & 73.4 \\
\bottomrule
\end{tabular}
\caption{Combined-attack robustness: ungated exact-message accuracy and
threshold-free detection AUROC. $N/N_{+}$ is 4{,}992/2{,}527 (AudioSeal),
5{,}008/2{,}537 (WavMark), and 5{,}008/3{,}552 (AURA); AUROC uses $N$ and
exact-message accuracy uses $N_{+}$.}
\label{tab:robust}
\end{table}

As a secondary check, validation-calibrated detect-then-decode accuracy over
watermarked and clean clips also improves in all six regimes for AudioSeal
($+0.9$--$+2.2$\,pp) and WavMark ($+2.0$--$+5.6$\,pp); full results and
protocol details are provided in the supplementary material.

\subsection{Capacity: Recovering Two Multiplexed Messages}

\begin{table*}[t]
\centering
\small
\setlength{\tabcolsep}{2.5pt}
\begin{tabular}{@{}llrrlrrl@{}}
\toprule
& & \multicolumn{3}{c}{Full coverage} &
\multicolumn{3}{c}{Partial coverage} \\
\cmidrule(lr){3-5}\cmidrule(lr){6-8}
System & Attacks & Baseline & Conf. & $\Delta$ [95\% CI] &
Baseline & Conf. & $\Delta$ [95\% CI] \\
\midrule
AURA & $K{=}1$ & 67.4 & 77.1 & $+9.7^{*}$ [$+8.0,+11.4$] &
14.7 & 62.7 & $+48.0^{*}$ [$+45.2,+50.8$] \\
& $K{=}2$ & 44.3 & 56.4 & $+12.1^{*}$ [$+10.3,+13.9$] &
9.7 & 42.1 & $+32.4^{*}$ [$+29.7,+35.0$] \\
& $K{=}3$ & 23.1 & 33.5 & $+10.4^{*}$ [$+8.7,+12.2$] &
4.6 & 23.0 & $+18.4^{*}$ [$+16.2,+20.7$] \\
\midrule
AudioSeal & $K{=}1$ & 0.1 & 16.8 & $+16.7^{*}$ [$+14.8,+18.6$] &
1.0 & 18.3 & $+17.3^{*}$ [$+15.1,+19.4$] \\
& $K{=}2$ & 0.1 & 12.8 & $+12.6^{*}$ [$+11.0,+14.4$] &
0.6 & 13.3 & $+12.7^{*}$ [$+10.8,+14.5$] \\
& $K{=}3$ & 0.2 & \phantom{0}9.8 & $+9.6^{*}$ [$+8.2,+11.2$] &
0.3 & \phantom{0}7.2 & $+6.9^{*}$ [$+5.5,+8.4$] \\
\midrule
WavMark & $K{=}1$ & 73.7 & \phantom{0}81.9 & $+8.2^{*}$ [$+6.2,+10.2$] &
64.7 & 74.5 & $+9.8^{*}$ [$+7.0,+12.6$] \\
& $K{=}2$ & 49.7 & \phantom{0}62.4 & $+12.7^{*}$ [$+10.3,+15.1$] &
39.0 & 51.3 & $+12.3^{*}$ [$+9.8,+15.0$] \\
& $K{=}3$ & 31.4 & \phantom{0}45.6 & $+14.2^{*}$ [$+11.6,+16.7$] &
22.1 & 33.0 & $+10.9^{*}$ [$+8.4,+13.4$] \\
\bottomrule
\end{tabular}
\caption{Controlled $ABAB$ multiplex recovery (\%). AURA is scored on two
raw 32-bit words (no BCH), AudioSeal and WavMark on two 16-bit payloads; only
within-system differences are interpreted. $\Delta$ is computed before
rounding; $^{*}$: Holm-adjusted $p<0.05$ over 18 comparisons.}
\label{tab:multiplex-controlled}
\end{table*}

  Table~\ref{tab:multiplex-controlled} shows that the Conformer improves
  joint-exact recovery in all 18 attacked conditions, with every difference
  remaining significant after joint Holm--Bonferroni correction. Because the
  systems differ in payload width, detector geometry, and source data, we
  interpret the within-system patterns of improvement rather than their absolute
  accuracies.

  For AURA, the gain under full coverage remains relatively stable across attack
  depth, at $+9.7$--$+12.1$\,pp, but rises to $+48.0$\,pp under partial coverage
  at $K{=}1$. The hard-vote baseline thresholds window outputs before
  aggregation and cannot identify which windows retain useful watermark
  evidence. The Conformer instead aggregates the soft sequence and can assign
  greater weight to informative windows without observing the coverage
  location. As $K$ increases, this gap decreases from $+48.0$ to
  $+18.4$\,pp while both methods deteriorate, suggesting a transition from an
  aggregation-limited regime to one in which the attacks increasingly destroy
  the underlying evidence.

  WavMark exhibits a different failure mode. Its grid-aware baseline reaches
  $99.6$--$100\%$ joint recovery on unattacked watermarked clips, confirming
  that its independent 1-s chunks contain sufficient information to recover both
  payloads when the nominal grid is known. Under attack, however, the released
  decoder retains only chunks whose 16-bit synchronization pattern matches
  exactly; a synchronization-bit error can therefore discard an otherwise
  informative payload estimate. The Conformer consumes the complete soft
  pattern-and-payload sequence and can combine such sub-threshold evidence.
  Accordingly, its full-coverage gain grows from $+8.2$\,pp at $K{=}1$ to
  $+14.2$\,pp at $K{=}3$, while the partial-coverage gains remain substantial
  throughout. The gain remains significant when all partial-coverage clips are included and
  clips lacking usable evidence for either payload are counted as failures.
  Thus, the improvement is not an artifact of filtering the primary evaluation
  to feasible clips.

  AudioSeal is more strongly constrained by the frozen detector itself. Its
  parity-wise baseline remains near zero in every attacked condition. On
  unattacked watermarked clips, a diagnostic continuous detector pass recovers
  $A$ exactly on $98.7\%$ of clips but $B$ on only $0.3\%$, indicating that the
  second payload is largely lost in the detector's temporal context before
  aggregation. The Conformer raises clean joint recovery to $19.3\%$ and
  improves every attacked cell, showing that some second-payload evidence
  remains recoverable. However, its decreasing gains as $K$ increases and its
  low absolute recovery indicate an upstream capacity limit that decoder-side
  aggregation alone cannot remove.

  Together, these results distinguish three decoder-side regimes: AURA benefits
  primarily from selecting localized soft evidence and WavMark from retaining
  evidence rejected by exact pattern gating, while AudioSeal is limited by a
  stronger representation bottleneck in the frozen detector.

  Alternating-slot embedding also costs no fidelity: over 512 unattacked
  WavMark clips, mean SNR is $42.57$\,dB for multiplex versus $42.59$\,dB for
  single-message embedding, with matching SI-SNR.

\subsection{Ablation Study}

  \begin{table}[t]
  \centering
  \small
  \setlength{\tabcolsep}{3.2pt}
  \begin{tabular}{@{}lrrrr@{}}
  \toprule
  Variant
  & \multicolumn{1}{c}{All}
  & \multicolumn{1}{c}{Partial}
  & \multicolumn{1}{c}{$K=2$}
  & \multicolumn{1}{c}{$K=3$} \\
  & \multicolumn{1}{c}{attacked}
  & \multicolumn{1}{c}{attacked}
  & \multicolumn{2}{c}{\small(full + partial)} \\
  \midrule
  Conformer & 58.52 & 52.66 & 57.34 & 39.75 \\
  BiGRU     & 53.86 & 48.46 & 51.51 & 34.23 \\
  Conf.--BiGRU      & $+4.67$ & $+4.20$ & $+5.83$ & $+5.52$ \\
  \midrule
  $-$ window-aux      & 58.14 & 52.45 & 56.69 & 39.24 \\
  $-$ slot flag       & 58.36 & 52.87 & 57.19 & 39.68 \\
  $-$ partial-cov aug & 57.41 & 51.22 & 55.61 & 39.46 \\
  \bottomrule
  \end{tabular}
  \caption{WavMark multiplex ablations: ungated joint-exact recovery (\%) on
feasible held-out positives, pooled over attacked conditions as defined in the
text. BiGRU replaces only the Conformer trunk. One training seed.}
  \label{tab:ablation}
  \end{table}
  We retrain three WavMark multiplex variants, removing one component at a time:
  the per-window auxiliary loss, the slot-flag input, or partial-coverage
  augmentation. WavMark provides a
  particularly informative setting because its grid-aware baseline is nearly
  perfect on unattacked watermarked clips, confirming that the frozen detector
  preserves sufficient evidence for both payloads. Under attack, its joint
  recovery remains in an intermediate range, avoiding the severe detector
  ceiling observed for AudioSeal. Moreover, WavMark's independent 1-s chunks
  and fixed 2-s slots provide a clear structure for studying all three
  components.
  
  The no-partial-coverage variant is retrained with partial-coverage augmentation
  disabled, so all positive training examples use full coverage. Its evaluation
  still includes the same full- and partial-coverage test conditions as the
  complete model.

All other settings are held fixed.

  Table~\ref{tab:ablation} reports ungated joint-exact recovery on paired
  held-out clips. The primary \emph{all-attacked} column pools the six full- and
  partial-coverage conditions at $K=1,2,3$, while the
  \emph{partial-attacked} column pools only the three partial-coverage
  conditions. Because the same source clips recur across conditions, confidence
  intervals use a source-clustered bootstrap and $p$-values use a source-level
  sign-flip test, with Holm correction across the three ablations.

  Only removing partial-coverage augmentation produces a detectable change,
  reducing all-attacked recovery by $1.11$\,pp
  (95\% CI $[0.44,1.80]$, Holm-adjusted $p=.006$). The reduction is larger in
  the partial-attacked subgroup ($1.44$\,pp, 95\% CI $[0.37,2.52]$,
  Holm-adjusted $p=.034$), consistent with the augmentation's intended role.
  We detect no difference after removing the auxiliary loss or slot flag.

  The no-slot variant omits the explicit $A/B$ flag during both training and
  inference but retains the dual pooling queries, fixed phase-aligned $ABAB$
  layout, and slot-gated auxiliary supervision. Its result therefore only suggests
  that the explicit flag is redundant under this specific synchronized protocol, not
  that the decoder can recover payloads under an unknown or shifted grid.

    \paragraph{Learned Aggregation Control} We replace the Conformer trunk with a
  small (171k) BiGRU while retaining the same inputs, window encoder,
  dual-query pooling, and output heads as a control for learned aggregation. The BiGRU recovers much of the benefit
  of learned aggregation, but the Conformer retains a consistent advantage (Table \ref{tab:ablation}),
  which is larger under chained attacks ($K{=}2,3$). This suggests that learning the
  temporal aggregation is the primary mechanism, while the higher-capacity
  Conformer provides additional robustness under heavier corruption.


    These ablations are specific to WavMark and do not establish that the same
  components have equivalent effects on the other watermarkers.

\section{Limitations}
Results use one training seed per system and predominantly Emilia speech, with
only a small FSD50K component; generalization to music, other datasets, and
other encoders remains untested.

For AudioSeal and WavMark, training and evaluation attacks come from the same
timing-warp-free family, so the reported robustness is in-distribution and
does not establish resistance to desynchronization. AURA's per-attack timing-warp
rows are out-of-distribution and may misalign the audio with its detection
windows. More generally, multiplex experiments assume a known,
sample-aligned $ABAB$ grid and do not test blind phase recovery after cropping
or temporal offsets.

Finally, multiplex results support only within-system comparisons because the
systems differ in payload width, sample rate, detector geometry, source data,
and training coverage. The fixed and learned decoders also follow their
intended system-specific temporal sampling paths rather than receiving
density-matched evidence. Thus, these experiments compare practical decoder
replacements rather than strictly evidence-matched aggregation architectures;
AudioSeal's absolute two-payload recovery also remains too low for a reliable
multiplex channel.

\section{Conclusion}
Time-collapsing aggregation limits both robustness and capacity, whether the
reduction averages, votes, or error-corrects a frozen detector's per-window
evidence. Replacing it with a small learned sequence decoder, while the encoder
and detector stay frozen, improves attacked message recovery and raises
detection AUROC on three distinct watermarkers. It also recovers two
time-multiplexed payloads that the shipped clip-wide decoders cannot separate,
with significant within-system gains over grid-aware fixed aggregation under
one to three chained attacks. AudioSeal's low absolute multiplex recovery shows
that aggregation cannot compensate when a frozen detector fails to preserve
both payloads. Reusing each watermarker through a thin per-window soft-output
adapter makes this a decoder upgrade rather than a redesign; natural next steps
are desynchronization attacks, non-speech audio, and more than two multiplexed
payloads.

\section{Generative AI Use Disclosure}
Large Language Models (LLMs) were used only for manuscript polishing, such as
rephrasing and grammar checks, and not for ideation, methodology, experimental
design, data analysis, or result interpretation. All scientific content was
produced and verified by the authors.


\bibliography{refs}

\end{document}

%% file: per_attack_tables.tex
\begin{table*}[t]
\centering
\small
\setlength{\tabcolsep}{2.2pt}
\begin{tabular}{@{}lrrlrrlrrl@{}}
\toprule
& \multicolumn{3}{c}{AudioSeal ($N_{+}=2{,}527$)}
& \multicolumn{3}{c}{WavMark ($N_{+}=2{,}537$)}
& \multicolumn{3}{c}{AURA ($N_{+}=3{,}552$)} \\
\cmidrule(lr){2-4}
\cmidrule(lr){5-7}
\cmidrule(l){8-10}
Attack
& Base. & Conf. & $\Delta$ [95\% CI]
& Base. & Conf. & $\Delta$ [95\% CI]
& Base. & Conf. & $\Delta$ [95\% CI] \\
\midrule
White noise & 6.1 & 6.1 & $0.0$ [$-0.9$, $+0.8$] & 0.3 & 0.5 & $+0.2$ [$0.0$, $+0.4$] & 0.3 & 0.9 & $+0.5^{*}$ [$+0.3$, $+0.8$] \\
Pink noise & 100.0 & 99.9 & $0.0$ [$-0.1$, $0.0$] & 100.0 & 100.0 & $0.0$ [$0.0$, $0.0$] & 88.3 & 95.2 & $+6.9^{*}$ [$+5.9$, $+7.9$] \\
Low-pass & 6.8 & 30.0 & $+23.2^{*}$ [$+21.3$, $+25.0$] & 98.6 & 100.0 & $+1.4^{*}$ [$+0.9$, $+1.9$] & 65.7 & 71.6 & $+5.8^{*}$ [$+4.4$, $+7.2$] \\
High-pass & 99.9 & 99.9 & $0.0$ [$-0.1$, $0.0$] & 100.0 & 100.0 & $0.0$ [$0.0$, $0.0$] & 95.1 & 98.8 & $+3.7^{*}$ [$+3.0$, $+4.4$] \\
Smoothing & 99.8 & 99.8 & $0.0$ [$-0.2$, $+0.2$] & 84.7 & 92.9 & $+8.2^{*}$ [$+7.1$, $+9.3$] & 86.6 & 88.9 & $+2.3^{*}$ [$+1.2$, $+3.4$] \\
Resample & 100.0 & 99.9 & $0.0$ [$-0.1$, $0.0$] & 100.0 & 100.0 & $0.0$ [$0.0$, $0.0$] & 88.8 & 95.9 & $+7.1^{*}$ [$+6.1$, $+8.1$] \\
Amplitude & 99.9 & 99.8 & $0.0$ [$-0.1$, $0.0$] & 100.0 & 100.0 & $0.0$ [$0.0$, $0.0$] & 95.0 & 98.7 & $+3.7^{*}$ [$+3.0$, $+4.4$] \\
Quantization & 98.3 & 97.6 & $-0.7$ [$-1.2$, $-0.2$] & 79.3 & 90.4 & $+11.1^{*}$ [$+9.8$, $+12.5$] & 60.4 & 69.1 & $+8.8^{*}$ [$+7.5$, $+10.0$] \\
Sample dropout & 99.9 & 99.7 & $-0.2$ [$-0.5$, $0.0$] & 67.0 & 86.8 & $+19.8^{*}$ [$+18.1$, $+21.5$] & 77.8 & 86.4 & $+8.6^{*}$ [$+7.4$, $+9.8$] \\
STFT masking & 91.6 & 91.4 & $-0.2$ [$-0.5$, $+0.1$] & 97.4 & 97.4 & $0.0$ [$-0.6$, $+0.6$] & 73.9 & 79.4 & $+5.5^{*}$ [$+4.4$, $+6.5$] \\
Median filter & 99.8 & 99.8 & $0.0$ [$-0.2$, $+0.2$] & 58.8 & 87.7 & $+28.9^{*}$ [$+27.1$, $+30.7$] & 86.9 & 93.7 & $+6.8^{*}$ [$+5.8$, $+7.8$] \\
\addlinespace[2pt]
Time-stretch & \multicolumn{3}{c}{--} & \multicolumn{3}{c}{--} & 88.8 & 95.1 & $+6.4^{*}$ [$+5.5$, $+7.3$] \\
Speed & \multicolumn{3}{c}{--} & \multicolumn{3}{c}{--} & 87.2 & 95.9 & $+8.7^{*}$ [$+7.6$, $+9.8$] \\
\bottomrule
\end{tabular}
\caption{Per-attack full-coverage single-message recovery rates (\%) on known-positive clips. AudioSeal and WavMark recover one 16-bit payload; AURA recovers one raw 32-bit payload without BCH. All metrics are ungated. Base.\ is each system's fixed decoder (native for AudioSeal and WavMark, a BCH-free hard vote for AURA) and Conf.\ the Conformer; $\Delta=\mathrm{Conf.}-\mathrm{Base.}$ Dashes: attack not evaluated for the 16-kHz systems. $^{*}$ denotes Holm--Bonferroni-adjusted $p<0.05$ within each system's forced-attack family.}
\label{tab:per-attack-combined}
\end{table*}